\input ppltexa.sty
\input epsf

% This is CFFK;ACCEL 29, finished on Jan. 27, 1981.
% revised on Apr. 17, 1981, and June 23, 1981.
% To be processed with TEX 117, and ACCEL REFSD
\let\rjustline\rightline
\def\bigglp{\biggl(}\def\biggrp{\biggr)}
\def\coth{\mathop{\rm coth}\nolimits}

\hsize 5.5 in \hoffset=0.5in
\vsize 8 in \def\pagesize{8.5 in} \voffset=0in
\long\def\endpaper{\ifempty\refsfile\else\par\input\refsfile\fi\par\vfill\end}

\defrefsfile{accelref.sty}
\def\refstyle{B}\def\seccase{L}\def\secnumbering{A} % For Physica D
\sections
\qdef\intro \qdef\correl \qdef\numer \qdef\univers \qdef\concl
\endpreamble
\raggedbottom
\predisplaypenalty 0

%\ctrline{D R A F T}
\rjustline{PPPL--1752 (Mar.\ 1981, revised Jun.\ 1981)}
\rjustline{Physica, \vol{4D}(3), 425--438 (Mar.~1982)}
\vskip 0.2 in
\hbox{\titl EFFECT OF NOISE ON THE STANDARD MAPPING}
\vskip 20 pt
\hbox{\hskip20pt Charles F. F. KARNEY}
\vskip 2 pt
\hbox{\hskip20pt Plasma Physics Laboratory, Princeton University,}
\hbox{\hskip20pt Princeton, New Jersey 08544, USA}
\vskip 5 pt
\hbox{\hskip20pt Alexander B. RECHESTER}
\vskip 2 pt
\hbox{\hskip20pt Plasma Fusion Center, Massachusetts Institute of Technology,}
\hbox{\hskip20pt Cambridge, Massachusetts 02139, USA}
\vskip 5 pt
\hbox{\hskip20pt Roscoe B. WHITE}
\vskip 2 pt
\hbox{\hskip20pt Plasma Physics Laboratory, Princeton University,}
\hbox{\hskip20pt Princeton, New Jersey 08544, USA}

\abstract

The effect of a small amount of noise on the standard mapping is considered.
Whenever the standard mapping possesses accelerator modes (where the
action increases approximately linearly with time), the diffusion
coefficient contains a term proportional to the reciprocal of the
variance of the noise term.  At large values of the stochasticity
parameter, the accelerator modes exhibit a universal behavior.  As a
result the dependence of the diffusion coefficient on the stochasticity
parameter also shows some universal behavior.


\section \intro.  Introduction

The standard mapping,
$$v_t-v_{t-1}=-\epsilon\sin x_{t-1},\qquad
x_t-x_{t-1}=v_t,\eqn(\en\sm)$$
is an important model with which to study the
phenomenon of stochasticity \ref\chirikov.
Recently, Rechester and White \ref\rw\ calculated the diffusion
coefficient for this mapping with an added noise term
$\delta x_t$ in the equation for $x_t-x_{t-1}$.  The
random variable $\delta x_t$ is sampled out of a normal distribution
$f(\delta x_t;\sigma)$ of variance $\sigma$,
$$f(x;\sigma) = (2\pi\sigma)^{-1/2}\exp(-\half x^2/\sigma).$$
Noise has been introduced into the standard mapping by other authors
\ref{\froeschle, \chirikova}.  However, they have primarily considered
$\epsilon\lsapprox1$ where the standard mapping (without noise) has
large regular regions.  The importance of the noise is to allow a
trajectory to wander from the regular to the stochastic regions leading
to a large change in the quality of the motion even when the noise is
small.

We shall be concerned here with the effect of a small amount of noise
on (\sm) in the stochastic regime $\epsilon\grapprox1$.  It is not so
clear in this case how important noise will be.  The primary
effect of the noise is to make the motion ergodic.  ``Ergodic'' here
means that for a single realization of the noise, time and phase-space
averages will be equivalent.  (When taking averages, phase space is
defined by taking $x$ and $v$ modulo $2\pi$.)  In the stochastic regime,
there are often small regular regions present with $\sigma=0$.  Since
their total area is small, one might think that they have little effect
when noise is introduced.  However, if some of the regular regions are
``accelerator modes,'' which are stable regions in which the particles
are continually accelerated \ref\chirikov, the diffusion coefficient
may exhibit a $1/\sigma$ dependence for small $\sigma$.
We may easily see the origin of this
$1/\sigma$ dependence.  Ergodicity ensures that a fixed fraction of time
is spent in each accelerator mode.  As $\sigma$ is decreased,
the mode is visited less often while the duration of each visit is
proportionately increased.  (Only the noise can cause a particle to
enter or leave an accelerator mode.)  This leads to a diffusion
consisting of a few large steps.  Since the step size $\Delta v$ and the
duration of the step are both proportional to $1/\sigma$, the
contribution to the diffusion coefficient also scales as $1/\sigma$.  In
this paper, we shall confirm the $1/\sigma$ dependence for two
different types of noise.  However, we expect similar behavior for any
noise model.

The remainder of this paper is organized as follows:  In section \correl,
the results of ref.\ \rw\ are interpreted in terms of the correlation
function.  This allows us to extend the result of ref.\ \rw\ to include
other noise models and to determine the behavior of the diffusion
coefficient when $\sigma$ is small.  In particular, we predict the
$1/\sigma$ dependence when there are accelerator modes present.
Section \numer\ looks at the behavior of the diffusion coefficient
numerically.  The $1/\sigma$ dependence is confirmed and the dependence
on $\epsilon$ is found.  In section \univers, we examine the
$\epsilon$ dependence in more detail.  We find that the accelerator
modes, which give the $1/\sigma$ behavior, have a universal structure when
$\epsilon$ is large.  This means that the diffusion coefficient
exhibits a dependence on $\epsilon$ which has some universal
characteristics.  The results are summarized in section \concl.

\section \correl.  The correlation function

We shall examine the standard mapping (\sm) with two different types of noise.
With the first type, the standard mapping becomes
$$v_t-v_{t-1}=-\epsilon\sin x_{t-1}+\delta v_{t-1},\qquad
x_t-x_{t-1}=v_t+\delta x_t,\eqn(\en\smd)$$
where $\delta v_t$ and $\delta x_t$ are random variables sampled
from distributions $f(\delta v_t;\rho)$ and $f(\delta x_t;\sigma)$
respectively.  This is a simple generalization of the system treated in
ref.\ \rw\ so that the noise causes diffusion in both the
$v$ and $x$ directions.

The effect of the second type of noise is
described by
$$v_t-v_{t-1}=-\epsilon\sin x_{t-1},\quad z_t-z_{t-1}=\delta z_t,\quad
x_t-x_{t-1}=v_t-z_t+\delta x_t.\eqn(\en\sml)$$
Here $(\delta x_t, \delta z_t)$ is a random point chosen with a
distribution $g(\delta x_t,\delta z_t;\sigma)$ where
$$g(x,z;\sigma) =
\exp(-\sigma)\, \delta(x,z) +
[1-\exp(-\sigma)]/(2\pi)^2\, u(x,z),$$
$\delta(x,z)$ is the Dirac delta function, and $u(x,z)$ is a
uniform function equal to $1$ if $\abs x <\pi$ and $\abs z <\pi$ and
$0$ otherwise.  This noise term models the effect of large-angle scatterings.
The effect of $z$ in (\sml) is to provide an origin
shift to $v$ in the equation for $x_t-x_{t-1}$.  The solution to (\sml)
behaves as that of the noiseless standard mapping (\sm) as long as
$(\delta x_t,\delta z_t)$ is $(0,0)$.  After an average of
$1/\sigma$ (for small $\sigma$) iterations,
a large-angle collision takes place
which completely randomizes the particle's position in $x$ and $z$.

Equations (\smd) and (\sml) represent examples of two distinct types of
noise.  In (\smd) the noise is diffusive; this is the limit in which the
particle suffers frequent but small uncorrelated kicks.  On the other
hand, in the large-angle scattering model (\sml), the particle is
rarely kicked by the noise but the kicks are large.  Chirikov
\ref\chirikova\ has considered (\smd) when $\sigma=0$, although he
introduced the noise as a shift of the origin of $v$ rather than
directly into the $v$ equation.  [Thus Chirikov's model is given by
(\sml) with $\delta x_t=0$ and $\delta z_t$ sampled from
$f(\delta z_t;\rho)$.]

In the absence of noise, there exist small regular regions
from which an orbit is excluded (assuming it started outside such a
region).  Noise destroys these regular regions allowing a particle to
wander anywhere in phase space.  In addition, the types of noise we
consider in (\smd) and (\sml) maintain the area-preserving nature of
the standard mapping because at any given time step
they merely translate the phase space by some fixed amount.  From this
it follows that the motion is ergodic, i.e., that time averages can be
replaced by phase-space averages.

The velocity-space diffusion coefficient is defined by
$$D=\lim_{t\to\infty}{\ave{(v_t-v_0)^2}\over2t},\eqn(\en\diffa)$$
where the angle brackets denote an average over some appropriately chosen
ensemble.  An equivalent definition is \ref\stocii
$$D={1\over2}C_0+\sum_{\tau=1}^\infty C_\tau,\eqn(\en\diffb)$$
where
$$C_\tau=\ave{a_{t+\tau} a_t},$$
$a_t=v_{t+1}-v_t$ is the acceleration, and the average now includes an
average over $t$.  Because the motion is ergodic, we can replace the
time average by a phase-space average coupled with an average over all
realizations of the noise terms.  (In calculating phase-space averages we
use the periodicity of the mappings in the $x$ and $v$ directions so
that the averaging need only be done over a $2\pi\times2\pi$ square.
However, when defining $D$ and $a_t$, the periodicity in $v$ is not used.)
The result for $C_\tau$ will be independent
of the ensemble chosen so the ensemble average can be ignored.  Thus we
have
$$C_\tau=\left<\int_0^{2\pi}{dx_0\over2\pi}\int_0^{2\pi}{dv_0\over2\pi}
a_\tau a_0\right>,\eqn(\en\cdef)$$
where here the angle brackets mean an average over the distributions of
all the noise terms appearing in $a_\tau a_0$.  For instance, for
the noise term in (\smd) we have
$$\eqalign{\ave{\hbox{ }}&=
\int_{-\infty}^\infty f(\delta x_{\tau+1};\sigma)\,d\delta x_{\tau+1}
\int_{-\infty}^\infty f(\delta v_\tau;\rho)\,d\delta v_\tau \quad\cdots\cr
&\quad\cdots\quad
\int_{-\infty}^\infty f(\delta x_1;\sigma)\,d\delta x_1
\int_{-\infty}^\infty f(\delta v_0;\rho)\,d\delta v_0.\cr}$$
This operation is an identity in the limit $\sigma\to0$
and $\rho\to0$.

The first few $C_\tau $ are then found for (\smd) to be
$$\eqalign{C_0&=\half\epsilon^2+\rho,\qquad
C_1=0,\cr
C_2&=-\half\epsilon^2 J_2(\epsilon) \exp(-\sigma-\half\rho),\cr
C_3&=-\half\epsilon^2 J_1^2(\epsilon) \exp(-\sigma-\rho)
+\half\epsilon^2 J_3^2(\epsilon) \exp(-3\sigma-\rho).\cr}$$
Assuming that the sum in (\diffb) can be truncated at $\tau=3$, we have
$$D\approx\half\rho+\half\epsilon^2[\half
-J_2(\epsilon)\exp(-\sigma-\half\rho)
-J_1^2(\epsilon)\exp(-\sigma-\rho)
+J_3^2(\epsilon)\exp(-3\sigma-\rho)].$$
With $\rho\to0$ this agrees with the result obtained by Rechester and
White \ref\rw.  Similarly when $\sigma\to0$, $D-\half\rho$ is the diffusion
coefficient given by Cohen and Rowlands \ref\cohen\ for Chirikov's
noise model \ref\chirikova.  (We must subtract $\half\rho$ to account for
the different way in which noise is introduced by Chirikov.)  We note
that the expression in Ref.\ \cohen\ also includes a term
$\half\epsilon^2 J_2^2(\epsilon) \exp(-\rho)$
which is part of $C_4$.
We might expect the truncation of (\diffb) to be accurate when
$\epsilon$ greatly exceeds the stochasticity threshold, i.e.,
$\epsilon\grgr1$.  This question will be examined in more
detail below.  This approach shows that the oscillations in $D$
seen by Chirikov \ref\chirikov\ are due to short-term correlations in
the standard mapping and, contrary to his assertion, are not directly
caused by the presence of accelerator modes.

The same calculation may be made for (\sml).  Here we obtain
$$\eqalign{C_0&=\half\epsilon^2,\qquad
C_1=0,\cr
C_2&=-\half\epsilon^2 J_2(\epsilon) \exp(-2\sigma),\cr
C_3&=[-\half\epsilon^2 J_1^2(\epsilon)
+\half\epsilon^2 J_3^2(\epsilon)] \exp(-3\sigma),\cr
C_\tau &=\exp(-\sigma\tau)C_\tau (\sigma\to0).\cr}$$
This result for $C_\tau $ comes about because the probability
that at least one
large-angle scattering takes place between $a_0$ and $a_\tau $ 
is $1-\exp(-\sigma\tau)$.  If no scattering takes place then the mapping
is the same as the noiseless one.

A similar method for deriving the diffusion coefficient is given
by Cary et al. \ref\cary\ who define a more general correlation
function.  While this method is more complicated than that described
above, it does allow the calculation of other statistical properties
of the mapping.
In order to illustrate this method of computing the diffusion
coefficient for another mapping, we consider the mapping obtained for
the motion of an ion in a lower hybrid wave \ref\stocii.  The mapping
is
$$x_t-x_{t-1}=2\pi\delta-2\pi A\cos y_{t-1},\qquad
y_t-y_{t-1}=2\pi\delta+2\pi A\cos x_t$$
and we are interested in diffusion in the $v$ direction where
$v=\half(y-x)$.  We take $A$ to be much larger than the
stochasticity threshold $A\grgr\quarter$.  Then the motion is
approximately ergodic (even though there is no noise in this model).
Furthermore, we expect (subject to the restrictions to be explored later
in this section) that only a few terms in (\diffb) contribute to
$D$.  So we have
$$D\approx\pi^2A^2[\half+J_0(2\pi A)\cos(2\pi\delta)
-J_1^2(2\pi A)\sin^2(2\pi\delta)],$$
where the first term in the brackets is the contribution from $C_0$ and
the other terms come from $C_1$.  Antonsen and
Ott \ref\ott\ have also derived this result using the method of
paths in Fourier space \ref\rrw.

The question of the accuracy of discarding the terms for
$\tau\grapprox3$ in (\diffb) may most easily be addressed with the
noise model employed in (\sml) because we need only determine the
behavior of $C_\tau(\sigma\to0)$ for $\tau\grapprox3$.
When $\sigma=0$ and $\epsilon\grapprox1$, phase
space may be divided into two regions:  a large connected stochastic
region and those parts of phase space within islands.  A particle
starting in either region stays forever in that region.  It is useful to
write $C_\tau $ as the sum of $C_\tau^{st}$ and $C_\tau^{is}$ which are the
contributions to the integrals in (\cdef) due to the stochastic and
island regions respectively.

We shall assume that $C_\tau^{st}$ decays exponentially with increasing
$\tau$.  So far as we know, this has not been proved for the standard
mapping.  The numerical evidence is that $C_\tau^{st}$ decays quite rapidly
for small $\tau$ and large $\epsilon$.  The decay for larger $\tau$ is
difficult to measure accurately because the error in the measurements
of $C_\tau^{st}$ may exceed the value of $C_\tau^{st}$.  (Grebogi and
Kaufman \ref\greb\ have numerical evidence confirming the exponential
dependence for quite large $\tau$.  On the other hand, the work of
Channon and Lebowitz \ref\channon\ on the H\'enon map \ref\henon\
shows an algebraic decay of correlations for a stochastic orbit.  But
their results do not preclude an exponential decay for longer times than
they were able to measure.)

The contributions due to the island region may be evaluated quite
accurately because, within a given island, the time-averaged
acceleration is a constant $a_i$ ($i$ is a subscript labelling the
various islands).  The frequency of oscillation around an island is
typically of order unity.  Therefore $C_\tau^{is}$ consists of a mean part
(independent of $\tau$) equal to $Q=\sum_ia_i^2A_i/A_0$ plus a
part which oscillates with a frequency of about unity.  $A_i$
is the area of the $i$th island and $A_0=(2\pi)^2$ is the total area of
phase space.

We are now in a position to assess the contributions to $D$ of
$C_\tau^{st}$ and $C_\tau^{is}$ when $\sigma$ is finite.  The sum
over $C_\tau^{st}$ probably converges rapidly so that a truncation
at some fairly low $\tau$ is quite accurate.  Since, in that case,
only the terms for small $\tau$ contribute to $D$, a small amount of
noise has little effect on this contribution.  The oscillatory part of
$C_\tau^{is}$ may similarly be neglected when evaluating $D$ since its
sum when weighted by $\exp(-\sigma\tau)$ is on the order of $\sigma$.
The mean part of $C_\tau^{is}$, on the other hand, increases $D$ by
$$\eqalignno{
D_{is}&={1\over2}C_0^{is}+\sum_{\tau=1}^\infty C_\tau^{is}
\approx Q\bigglp{1\over2}+\sum_{\tau=1}^\infty\exp(-\sigma\tau)\biggrp\cr
&=\half Q\coth(\half\sigma)\approx Q/\sigma\cr
&=\sum_i{a_i^2A_i\over\sigma A_0}.&(\en\disl)\cr}$$
This will be nonzero if at least one of the islands is an
accelerator mode, i.e., $a_i\neq0$ for some $i$.  In these modes,
a particle, instead of returning to the original island after $N$
iterations, goes to the image of that island displaced upwards
or downwards in $v$ by some multiple of $2\pi$.  Such a mode is called an
$N$th-order accelerator mode.  (In fact, due to the
symmetries of the standard mapping, they come in pairs
with $A_{2i}=A_{2i-1}$ and $a_{2i}=-a_{2i-1}$.)  So, if accelerator
modes exist, $\sigma$ can be chosen so that $D_{is}$ and hence $D$ are
arbitrarily large.

Accelerator modes are best found by looking for stable accelerating
fixed points.  Around each such fixed point there will be an accelerator
mode.  Several first- and second-order accelerating fixed points for
(\sm) are cataloged in table \islands.  Chirikov \ref\chirikov\ gives
$$\epsilon_0=2\pi n < \abs\epsilon < 
[(2\pi n)^2+16]^{1/2}=\epsilon_1\eqn(\en\first)$$
with $n$ being an integer as the condition for the stability of
first-order fixed points.  The magnitude of the acceleration of the
accelerator mode associated with such fixed points is $2\pi n$.
When $\epsilon=\epsilon_1$, the first order fixed points
become hyperbolic with reflection and a pair of stable second-order
fixed points (marked by asterisks in the table)
are born.  The accelerating regions around these
second-order fixed points are best thought of as being a continuation
of the first-order accelerator mode.

The physical explanation of the $1/\sigma$ divergence was given in
section \intro.  Here we will cast that explanation into more
quantitative terms.  Consider a particle that has just been placed in an
accelerator mode by a collision.  When $\sigma$ is small, the
probability that it survives in that mode for longer than a time $t$ is
$P(t)=\exp(-\sigma t)$.  (This is just the probability that there is no
collision during the time $t$.)  The probability that it leaves between
times $t$ and $t+dt$ is $p(t)dt$ where
$p(t)=-dP(t)/dt=\sigma\exp(-\sigma t)$.  From (\diffa), the contribution
to the diffusion coefficient from the accelerator modes is
$$D_{is}=\sum_i{1\over2}
{\int_0^\infty a_i^2t^2p(t)dt\over\int_0^\infty tp(t)dt}
{A_i\over A_0}.\eqn(\en\disla)$$
The factor $A_i/A_0$ is the fraction of time a particle spends in the
$i$th accelerator mode.  Substituting for $p(t)$, we have
$D_{is}=Q/\sigma$ which agrees with (\disl).

The same considerations apply to the standard mapping with noise given
by (\smd).  In this case, the form of $p(t)$ is not known; it will in
fact depend on the size of the accelerator mode.  However, we do expect
the duration of a particle's stay in an accelerator mode to be
approximately $\Delta^2/\sigma$ (for $\rho=0$) where $\Delta$ is the
scale length of the island.  This is to be compared with an average
duration of $1/\sigma$ for the large-angle scattering case (\sml).
Since $\Delta$ is usually quite small, the coefficient of the
$1/\sigma$ term in $D$ for (\smd) should be smaller than that for
(\sml).  These considerations will be refined in section \univers, when
we will be able to make more accurate scaling arguments.

In the next section we numerically confirm the $1/\sigma$ dependence
and explore the dependence of $D$ on $\epsilon$.

\section \numer.  Numerical evaluation of the diffusion coefficient

In order to measure the diffusion coefficient numerically we adopted a
method based on (\diffb) which is designed to handle systems with long
correlations.  The trajectories of $J$ particles with random initial
conditions (with a uniform distribution)
are advanced to $t=kT$ according to either (\smd) or (\sml).
A correlation function is defined for the $k$th iterate of the map for
each trajectory by
$$C_\tau^k={1\over T-\tau}\sum_{t=0}^{T-\tau-1} a_t^ka_{t+\tau}^k$$
where $a_t^k=v_{k(t+1)}-v_{kt}$ is the acceleration due to $k$ iterates
of the map.  A diffusion coefficient based on the $j$th trajectory is
given by
$$D_j={1\over2k}C_0^k+{1\over k}\sum_{\tau=1}^L C_\tau^k.$$
The final value of $D$ is obtained by averaging $D_j$ over $j$.  The
standard deviation of $D_j$ divided by $\sqrt{J}$ is used to give a
measure of the error in $D$.

In this method, correlations up to a time separation of $kL$ are retained.
With $T=1$ and $L=0$, we recover the ``standard'' method which is
based on (\diffa); in order to obtain accurate results in this case $J$
must be large.  Here we do not take $J$ to be large; however, good
statistics are obtained because we take $T-L\grgr1$ so that there are
many observations of each $C_\tau^k$.  Normally we take $J$ to be 64
which allows us to make full use of the vectorization capabilities of
the Cray--1 computer on which the computations of $D$ are performed.

In fig.\ \dvssig, we show the $\sigma$ dependence of $D$ for (\smd)
with $\rho=0$ and $\epsilon=6.6$ and $12.8$.  $D$ is normalized to its
quasi-linear value $D_{ql}=\quarter\epsilon^2$.  These values of
$\epsilon$ were chosen to satisfy (\first) for $n=1$ and $2$.  We see
that $D$ does have a $1/\sigma$ dependence for small $\sigma$.  The
values of $\sigma$ at which this dependence becomes evident are about
$10^{-4}$ and $10^{-5}$ for $\epsilon=6.6$ and $12.8$.  The values for
$D$ for $\sigma=10^{-5}$ greatly exceed the numerical values given in
ref.\ \rw.  Since orbits of length $50$ were used in those computations,
the effect of the accelerator modes was largely suppressed.

Also shown in fig.\ \dvssig\ is $D$ for (\sml) with $\epsilon=6.6$.  As
expected, the coefficient of the $1/\sigma$ term is nearly $100$ times
larger than for (\smd).

Taking the
limit $D(\sigma\to0)$ gives an infinite result from the $1/\sigma$
term.  If we interchange the limits so that we take $\sigma\to0$ before
$t\to\infty$ in (\diffa), the value of $D$ depends on how the
initial conditions are chosen since the motion in this case is not
ergodic.  If an ensemble is defined by choosing
initial conditions uniformly in
phase space, $D$ is infinite because some trajectories will be
accelerating.  This is then consistent with the value of $D$
obtained by taking the limit $\sigma\to0$ after $t\to\infty$.  A more
``natural'' ensemble is obtained if we restrict the initial conditions
to the stochastic region of phase space.  Figure\ \dvssig\ shows the
value of the $D$ for $\epsilon=6.6$ and $\sigma=0$ with such initial
conditions.  The error in this measurement of $D$ is quite
large even though long trajectories were used in the computation.  This
probably arises because the stochastic region includes a ``sticky''
portion close to the accelerator modes.  A particle which wanders into
this portion of phase space may still spend a long time accelerating
even though the trajectory is still stochastic.  The properties of
these sticky regions around islands need more thorough study if the
diffusion coefficient for $\sigma=0$ is to be understood.

No enhancement of $D$ was detected in the ranges of $\epsilon$ where
second-order accelerator modes exist.  Because these modes are much
smaller than the first-order accelerator modes the value of $\sigma$ at
which they begin to contribute significantly is so small that
prohibitively long runs would have to be made to detect any effect
numerically.

Next we turn to the behavior of $D$ as a function of $\epsilon$.  Here
we hold $\sigma$ fixed and equal to $3\times10^{-6}$ and $\epsilon$ is
varied in and somewhat beyond the ranges given by (\first) with $n=1$ and
$2$.  The results are shown in fig.\ \dvseps.  $D$ rises quite rapidly
as soon as $\epsilon$ exceeds $\epsilon_0$ for the first-order fixed
point.  At about one quarter and at about one half of the way through the
interval $(\epsilon_0,\epsilon_1)$, $D$ is dramatically reduced.  As
we shall see this is due to the appearance of fourth- and third-order
resonances.  Nothing much happens to $D$ at $\epsilon_1$.  Although the
central fixed point becomes unstable at this value of $\epsilon$, there
is still a KAM surface of the original topology surrounding both the
unstable first-order fixed point and the two new second-order stable
fixed points.  There is little change in the overall size of
the island at this transition.

Perhaps the most noteworthy feature of fig.\ \dvseps\ is that the plots
for both $n=1$ and $2$ are so similar.   This suggests that there may
be a universal structure for the accelerator modes.  We pursue this
subject further in the next section.

\section \univers.  Universal behavior of accelerator modes

Referring to table \islands, we see that the accelerator modes exist
only in a narrow range in $\epsilon$.  They are likewise present only in
a small region of phase space.  This effect becomes more pronounced as
$\epsilon$ is increased and allows us to approximate the
accelerator modes by a Taylor-series expansion of the mapping.  We
consider a general area-preserving map of the $(x,y)$ plane which
depends on a parameter $k$.  We shift the origin and $k$ so that the
accelerator mode first appears at $(x,y)=0$ and $k=0$.  We pick a frame
traveling with the acceleration of the mode; therefore the
constant terms which represent the acceleration are subtracted.  We
shall only directly treat accelerator modes which appear as a result of
tangent bifurcations.  Other higher-order fixed points which come from
bifurcations of an existing accelerator mode will be treated
as part of that accelerator mode.
The linear terms of the mapping at $k=0$ have the form
\def\p{\prime}
$$x_1=(1+a^\p)x_0+by_0,\qquad y_1=b^\p x_0+(1-a^\p)y_0,$$
with $a^{\p2}+bb^\p=0$.  Because the mode appears as a tangent bifurcation,
the trace of the tangent mapping
matrix is $2$.  By transforming $(x,y)$ with
$$x=x^\p+a^\p y^\p/(a^{\p2}+b^2),\qquad y=-a^\p x^\p/b+by^\p/(a^{\p2}+b^2),$$
the linear mapping becomes
$$x_1^\p=x_0^\p+y_0^\p,\qquad y_1^\p=y_0^\p.$$
A similar transformation is possible if $b=0$ but $b^\p\neq0$.  The
only case where the transformation is not possible is if $a^\p=b=b^\p=0$
in which case the linear mapping is an identity.
We now add the terms in the Taylor expansion which are quadratic in $x$
and $y$ and linear in $k$,  
$$\eqalignno{
x_1^\p&=x_0^\p+y_0^\p+cx_0^{\p2}+dx_0^\p y_0^\p+ey_0^{\p2}+fk&(\e a)\cr
y_1^\p&=y_0^\p+c^\p x_0^{\p2}+d^\p x_0^\p y_0^\p+e^\p y_0^{\p2}+f^\p k.
&(\+ b)\cr}$$
We have taken $k\sim O(x^2)$.
Unfortunately, this mapping is not in general area-preserving because
the omitted cubic terms in the map also contribute to
area preservation.  However, we can make (\+) conserve area by expressing
four of the coefficients of the quadratic terms in terms of the other
two.  The coefficients of $x_0^{\p2}$, $c$ and $c^\p$, are taken as the
independent ones.  This choice is motivated by noting that the results
are then independent of our choice of the direction of time.  Thus when
we invert (\+), the coefficients of $x_1^{\p2}$ depend only on $c$ and
$c^\p$, whereas the other coefficients are linked in a more complicated
way.  Also, in some respects we may order $y^\p$ as $x^{\p2}$ [e.g.,
consider the positions of the fixed points of (\+)].  The
terms involving $x_0^\p y_0^\p$ and $y_0^{\p2}$ are of the same order
as the neglected terms.  Therefore their coefficients should only be
chosen to ensure the preservation of area; i.e.,
$$d^\p=2(c^\p-c),\quad 
e^\p=(c^\p-c)^2/c^\p,\quad d/d^\p=e/e^\p=c/c^\p.$$
Equation (\+) may then be written as
$$\eqalign{
x_1^\p&=x_0^\p+y_0^\p+c[c^\p x^\p+(c^\p-c)y^\p]^2/c^{\p2}+fk,\cr
y_1^\p&=y_0^\p+[c^\p x^\p+(c^\p-c)y^\p]^2/c^\p+f^\p k.\cr}$$
Finally, we perform the transformations
\def\pp{^{\prime\prime}}
$$x^\p=2[x\pp-(1-c/c^\p)y\pp]/c^\p,\qquad y^\p=2y\pp/c^\p$$
and
$$x\pp=X,\qquad y\pp=Y+2K(f/f^\p-c/c^\p),\qquad k=-4K/(c^\p f^\p)$$
to give
$$Y_1-Y_0=2(X_0^2-K),\qquad X_1-X_0=Y_1.\eqn(\en\uni)$$
This is a universal mapping approximating the behavior of accelerating
modes for large stochasticity parameter.
All the transformations which give (\+) are linear, and,
with the exception of the last one which is just a shift of
the origin, they are all independent of $k$ (and $K$).
For the first-order accelerator modes for the standard mapping (\sm)
which appear at $\epsilon=2\pi n$, the transformations reduce to
$$x=\pm\bigglp{\pi\over2}+{2\over\pi n}X\biggrp,
\quad v=\pm\bigglp-2\pi nt+{2\over\pi n}Y\biggrp,
\quad\epsilon=2\pi n+{4\over\pi n}K.\eqn(\en\firsttx)$$

The transformations leading to (\uni) are not well defined if the
various coeffi\-cients satisfy unusual relationships.  The first
transformation is not possible if the linear term is an identity.
If the motion around a stable $N$th-order fixed point of a mapping is
similar to rotation by an angle $2\pi p/q$ where $p$ and $q$ are
integers, then the linear part of the $qN$th iterate of the map is an
identity.  But this does not correspond to the first appearance of an
accelerator mode.  (It appeared with the $N$th-order fixed point or
sooner.)  The transformations also fail if
$c^\p$ or $f^\p$ is zero.  In
this case higher order terms have to be kept.  This is what happens
with the standard mapping at its (non-accelerating) fixed point which
appears at $\epsilon=0$ and $(x,v)=(0,0)$.  The same thing happens if
we look at the second-order fixed points which appear when a
first-order fixed point goes unstable (the fixed points marked with
asterisks in table \islands).  But here again such second-order fixed
points are not the first occurrence of an accelerator mode in the
neighborhood of parameter and phase space.  They are treated by the
second-order fixed points of (\uni) that appear for $K>1$.  Another
form of degeneracy occurs if the exact mapping has, for instance, a
real square root.  Then quadratic terms in the mapping may be finite
but still the cubic and quartic terms may not be neglected.  An example
of this is provided by the second iterate of (\uni).

Figure \unicheck\ illustrates this mapping for a particular value of
$K$ and also shows that it does indeed closely approximate the standard
mapping near first- and second-order accelerator modes.  Thus
accelerator modes may be studied by examining (\uni) and how the
transformations affect the diffusion.

We begin by cataloging some properties of (\uni).  For $0<K<1$, (\uni)
may be transformed into the H\'enon quadratic map \ref\henon,
$$x_1=x_0\cos\alpha-(y_0-x_0^2)\sin\alpha,\quad
y_1=x_0\sin\alpha+(y_0-x_0^2)\cos\alpha.$$
$K$ is related to H\'enon's parameter by
$K=\sin^4(\alpha/2)$.  The transformation between the two sets of
coordinates is given by
$$X=\cos^2(\alpha/2)\sin(\alpha/2)w+Y/2-\sin^2(\alpha/2),\quad
Y=-2\cos(\alpha/2)\sin^2(\alpha/2)v,$$
and
$$w=x\cos(\alpha/2)+y\sin(\alpha/2),\quad
v=-x\sin(\alpha/2)+y\cos(\alpha/2).$$
The transformation depends intimately on $\alpha$ (and hence on $K$)
and becomes singular at $K=0$ and $1$.

For $K<0$, (\uni) has no fixed points.  It
has two first-order fixed points for $K>0$,
$(X,Y)=(\mp\sqrt K,0)$, which following ref.\ \henon\ we call $I_1$ and
$I_1^\p$ respectively.  $I_1^\p$ is always hyperbolic and so is unstable.
$I_1$ is elliptic and therefore stable for $K<1$.  At $K=1$, $I_1$ turns
into a hyperbolic point with reflection and two second-order fixed
points are born.  Thus the values $K=0$ and $K=1$ correspond to the
parameter values $\epsilon_0$ and $\epsilon_1$ listed for the fixed
points without asterisks in table \islands.

The second-order fixed points are stable for $1<K<\fract5/4$.  This
corresponds to the range in $\epsilon$ for which the fixed points
labelled by asterisks in table \islands\ are stable.    At
$K=\fract5/4$, a second bifurcation takes place giving rise to periodic
trajectory of period 4.  This trajectory in turn becomes unstable at
$K=1.2801$ when a period-8 cycle is born.  The process of a period-$2^m$
trajectory becoming unstable and producing a period-$2^{m+1}$
trajectory continues.  Greene et al. \ref\mackay\ show that it
accumulates at $K=1.2840=\quarter(1+1.2663)^2$.  When $K$ exceeds this
value, they conjecture that (\uni) has no stable fixed points.
Therefore, for large $n$ the first-order accelerator modes exist for
$$\epsilon_0=2\pi n<\epsilon<\epsilon_2\approx
\epsilon_0+1.2840(\epsilon_1-\epsilon_0)\approx
2\pi n+1.2840\times4/(\pi n).\eqn(\en\firsta)$$

The map (\uni) has two main symmetry lines.  They are the $X$-axis
($Y=0$) which corresponds to the $w$ axis in the H\'enon map and the
line $Y=K-X^2$.  Reflection in one of the symmetry lines corresponds to
reversing time.  E.g., $Y=0$ is the invariant line for the
transformation $X=X^\p-Y^\p$, $Y=-Y^\p$ which turns (\uni) into its
inverse.

In order to understand how (\uni) contributes to diffusion we must
determine how large a region is trapped around the stable fixed
points.  We define an orbit to be trapped if $X$ and $Y$ remain bounded
for all time.  (Trapped orbits are the ones which contribute strongly to
the diffusion because in the original mapping they are the ones that
are perpetually accelerated.)  Untrapped orbits escape to infinity with
$Y\to+\infty$.  It is straightforward to show that for $K\leq2$ all
particles with $Y\geq10$ escape in this way.  An approximate
numerical test for being trapped is to check that $Y<10$ during a large
number of iterations.  Figure \trap\ shows the extent of the trapped
region along the two symmetry lines and its total
area as functions of $K$.  Figure \trap b is just a rescaling of
H\'enon's fig.\ 2.  The dips in the plots of the diffusion
coefficient (fig.\ \dvseps)
correspond to the large reductions in the size of the
trapped region that occur for certain values of $K$.
These are associated with the occurrence of higher-order
resonances to the basic rotation about the $I_1$ \ref\henon.  Particularly large
effects are produced by the fourth- and third-order resonances.  A
stable period-four cycle exists for $\quarter<K<0.3044$.  This is born
at $I_1$ and in the process of moving away from this point destroys
much of the stable region.  A stable period-three cycle $I_3$ exists for
$\half<K<\fract9/{16}=0.5625$.  Unlike other resonances this is born
away from the associated lower-order fixed point $I_1$.  $I_3$ moves
further away from $I_1$ as $K$ increases, while its unstable twin
$I_3^\p$ moves towards $I_1$.  When $I_3$ goes unstable at
$K=\fract9/{16}$, $I_3^\p$ is at $I_1$ causing the destruction of all
stable regions in the neighborhood of $I_1$.  These two resonances
cause the large reductions in $D$ seen in fig.\ \dvseps.  Thus, the
values of $\epsilon$ at which the dips occur are given by
$K\approx0.3044$ and $0.5625$ (with $n=1$ and $2$) in (\firsttx).

In order to make a more quantitative comparison between the behavior
of the diffusion coefficient in fig.\ \dvseps\ and the behavior of
(\uni), we must ascertain the effect of the transformations used to
derive (\uni).  We saw in section \correl\ that the contribution of an
accelerator mode to the diffusion coefficient when the noise is due to
large-angle scattering (\sml) is given by the area of the mode and
its acceleration.  Now the area of the accelerator mode is $4/(\abs b
c^{\p2})$ of the area of the trapped region of (\uni) (fig.\ \trap c).
So in the limit of small $\sigma$, the diffusion coefficient for
(\sml) consists of a superposition over accelerator modes
of forms of fig.\ \trap c, linearly
scaled by appropriate amounts in both directions.
For a first-order accelerator mode, $A_i$ the area of the mode scales as
$1/n^2\sim 1/\epsilon^2$ (\firsttx) while the acceleration varies
as $a_i\sim n\sim \epsilon$.  Thus from (\disl) we have $D_{is}\sim 1/\sigma$.  The
relative importance of the first-order modes with this noise model
is given by $D_{is}/D_{ql}\sim 1/(\sigma\epsilon^2)$.

With the noise employed in (\smd), the situation is more complicated
because the way diffusion due to $\sigma$ and $\rho$ enters (\uni)
depends in a more involved way on the transformations used to derive
(\uni).  In particular, although fig.\ \trap c has the same qualitative
features as fig.\ \dvseps, the diffusion coefficient is no longer
simply a function of the area of the accelerator mode.
For instance, take the case
where $\rho=0$ (then all the diffusion due to the noise is in the
$x$-direction).  In (\uni) the diffusion is then along a line
$Y=a^\p(a^{\p2}+b^2)X/b^2$.  So the effect of each accelerator mode as a
function of $K$ depends on what this direction is.  However, in (\smd)
the important accelerator
modes are the first-order ones and for these modes the
transformation to $(X,Y)$ takes the particularly simple form
(\firsttx).  The only way the transformation changes as $n$ is changed
is by an overall scale factor.  The effect of the noise
will then only depend on the ratio of $\rho$ to $\sigma$ (apart from a
constant factor).

In order to find the effect of a first-order accelerator mode in (\smd)
on the diffusion coefficient, we go back to the expression for $D_{is}$
given in (\disla).  Now $p(t)$ is no longer a probability since it can
not be normalized.  It may be defined as follows.  Imagine
starting infinitely many particles at the boundary of the trapped
region.  (The number of particles has to be infinite because almost all
of them leave the trapped region immediately.)  Suppose the number which
are left after a time $t$ is $P(t)$.  The number which leave between
$t$ and $t+dt$ is then $p(t)dt$ where $p(t)=-dP(t)/dt$.  Before
applying (\disla), we must also ensure that $\sigma$ and $\rho$ are
small so that each visit to the accelerator mode is uncorrelated with
the previous one.  (If $\sigma$ or $\rho$ is not small, the particle may
be immediately scattered back into the trapped region after being
scattered out of it, instead of being swept far away from the island.
This effectively increases the area of the trapped region.  Such
considerations were not necessary for the large-angle scattering
model.)  Taking $\rho=0$, we have $p(t)\sim q(\sigma t/\Delta^2)$ where $q$
is some function which, for a given $K$, applies for all first-order
accelerator modes and which is independent of $\sigma$.  $\Delta$ is the
scale length of the island which from (\firsttx) is proportional to
$1/n$.  Here again, $\sigma$ must be small to be able to write $p(t)$ in
this way because we need to be able to separate the slow time scale of
the motion due to the noise in (\smd) from the fast time scale due to
the standard mapping itself.  This form for $p(t)$ applies for any
diffusive noise model if we regard $\sigma$ as a measure of the extent of the
Green's function response after a unit of time under the action of the
noise alone.  Using (\disla), we find that, when $\epsilon$ satisfies
(\firsta), $D_{is}\sim 1/(\sigma\epsilon^2)$ and
$D_{is}/D_{ql}\sim 1/(\sigma\epsilon^4)$.  This is confirmed by fig.\ 
\dvseps, where we see that $D/D_{ql}$ decreases by roughly a factor of
$16$ when $\epsilon$ is doubled.

\section \concl.  Conclusions

When noise is added to the standard mapping the diffusion coefficient
consists of two parts.  One part is primarily due to short-term
correlations in the stochastic region of phase space and this part
exhibits nearly sinusoidal oscillations \ref\rw.
The other part is proportional
to the inverse of the noise parameter $\sigma$ and exists only when
accelerator modes are present.  When the noise is weak, this latter part
dominates.  The role of the noise is to ensure that every trajectory
eventually visits the accelerator modes.  This effect has been
confirmed numerically for the first-order accelerator modes.  These
modes exist in windows given by (\firsta) and their relative importance
decays as $\epsilon^{-4}$ for diffusive noise (\smd) and as
$\epsilon^{-2}$ for large-angle scattering (\sml).
Although many higher-order accelerator modes
exist (see table \islands), they are so small that they do not
contribute significantly to $D$ for the values of $\sigma$ that can be
dealt with computationally.  In the absence of noise, the diffusion
coefficient is not fully defined until the ensemble in (\diffa) is
specified.  However, anomalies in the diffusion coefficient are seen
even when the ensemble in (\diffa) is taken to be trajectories in the
stochastic region.  This is apparently because trajectories can be
greatly accelerated while in the vicinity of the accelerator modes.

We showed in section \univers\ that the accelerator modes normally exhibit a
universal behavior.  De\-pen\-ding on the noise model, the diffusion
coefficient therefore exhibits some universal features as the parameter
$\epsilon$ is varied.

An interesting accelerator mode which is not accurately described by
(\uni) is found near $\epsilon=3$.  At this value of the parameter,
Chirikov \ref\chirikov\ noticed that the distribution of particles
after $100$ iterations of the standard mapping was characterized by two
diffusion rates.  (See fig.\ 5.8 of ref.\ \chirikov.)  This is due to the
existence of a 4th order accelerator mode with acceleration $\abs
a/(2\pi)= \quarter$.  The history of this mode is somewhat unusual.  The
mode is born as a tangent bifurcation at $\epsilon = 2.9453$.  The
period-4 orbit becomes unstable at $\epsilon = 2.9775$, giving rise to
a pair of period-8 orbits.  However, these period-8 orbits do not become
unstable as $\epsilon$ is increased.  Rather they are reabsorbed into
the period-4 orbit making it stable again (this happens at $\epsilon =
2.9950$).  At $\epsilon = 3.1068$, it becomes unstable as an ordinary
hyperbolic orbit.  Simultaneously, two stable period-4 orbits are created.
They become unstable at $\epsilon = 3.1449$ giving rise to twice as
many period-8 orbits.  The standard period-doubling process now takes
place as $\epsilon$ is increased.  (So actually the accelerator mode
exists for $\epsilon$ somewhat larger than $3.1449$.)  This degenerate
behavior is probably due to the fact that this accelerator mode lies on
one of the symmetry lines of the standard mapping, $v=n\pi+\half\epsilon
\sin(x)$ with $n$ being an integer.

Similar effects are expected in other mappings which allow accelerator
modes.  Such mappings are ones which are periodic in the velocity
direction.  For example, an anomalously large diffusion coefficient was
found at a certain value of the parameter for the mapping studied in
ref.\ \stocii.  (The divergent behavior of the diffusion coefficient for
this case has been noted by Antonsen and Ott \ref\ott.)
However, such mappings are often derived as
approximations to mappings which are not periodic in the velocity
direction.  For instance (\sm) may be derived from similar mappings in
which $\epsilon$ is effectively a function of $v$, $\epsilon(v)$
\ref\llc.  These more general mappings do not have accelerator modes.
Nevertheless, if the dependence of $\epsilon$ on $v$ is weak, we still
expect there to be some effect due to
the presence of the ``latent'' accelerator modes.
These are channels in phase space which allow particles to be
uniformly accelerated from a velocity $v_0$ such that
$\epsilon(v_0)\approx\epsilon_0$ to a velocity $v_2$ such that
$\epsilon(v_2)\approx\epsilon_2$ ($\epsilon_0$ and $\epsilon_2$ being
the parameter values for the birth and death of an accelerator mode in
the standard mapping).  No noise is required to bring particles into
these channels.  This short cut will result in an enhancement of the
diffusion coefficient between $v_0$ and $v_2$.  Diffusion on longer
velocity-space scales will still be limited by the slower diffusion
elsewhere.

Finally, we note that the $1/\sigma$ dependence of $D$ may arise in the
study of particle confinement in fusion devices,
whenever a small group of particles is unconfined in the collisionless
limit.  (For problems involving diffusion in real space, the
unconfined particles, for which the position coordinate is
monotonically increasing, correspond to accelerator modes.)
Collisions, which can scatter these particles into confined
orbits, serve to slow the loss of these particles leading to a
diffusion process which is inversely proportional to the collision
frequency \ref\connor.

\acknowledgments

This work was supported by the U.S.  Department of Energy under
Contracts DE--AC02--76--CH03073 and DE--AS02--78ET53074.  This work was
begun while two of us (C.F.F.K. and A.B.R.) were attending the Aspen
Institute of Physics.  We would like to thank H. A. Rose whose valuable
comments stimulated us to look into this problem and R. S. MacKay and
P. J. Morrison for useful discussions.

\endpaper